\documentclass[twocolumn,showpacs,preprintnumbers,amsmath,amssymb]{revtex4}

\newcommand{\commentoutA}[1]{}

\usepackage{graphicx}
\usepackage{dcolumn}
\usepackage{bm}
\begin{document}

\preprint{LA-UR 06-2984}

\title{Time-reversible Born-Oppenheimer molecular dynamics}

\author{Anders M. N. Niklasson \footnote{Corresponding author: Email amn@lanl.gov}}
\affiliation{Theoretical Division, Los Alamos National Laboratory, Los Alamos, New Mexico 87545}
\affiliation{Applied Materials Physics, Department of Materials Science and
Engineering, Royal Institute of Technology, SE-100 44 Stockholm, Sweden}
\author{C. J. Tymczak, and Matt Challacombe} 
\affiliation{Theoretical Division, Los Alamos National Laboratory, Los Alamos, New Mexico 87545}

\date{\today}

\begin{abstract}
We present a time-reversible Born-Oppenheimer molecular dynamics scheme, based on
self-consistent Hartree-Fock or density functional theory, where both the
nuclear and the electronic degrees of freedom are propagated in time.
We show how a time-reversible adiabatic propagation of the electronic degrees 
of freedom is possible despite the non-linearity and incompleteness of the self-consistent 
field procedure.  Time-reversal symmetry excludes a systematic long-term energy 
drift for a microcanonical ensemble and the number of self-consistency
cycles can be kept low (often only 2-4 cycles per nuclear time step) thanks to a good initial 
guess given by the adiabatic propagation of the electronic degrees of freedom.
The time-reversible Born-Oppenheimer molecular dynamics scheme therefore combines a low computational
cost with a physically correct time-reversible representation of the dynamics,
which preserves a detailed balance between propagation forwards and backwards
in time.  
\end{abstract}

\maketitle

{\it Ab initio} molecular dynamics based on Hartree-Fock or density functional
theory \cite{Roothaan,hohen,KohnSham65,RCar85,MCPayne92,DMarx00,HBSchlegel01,PPulay04,JMHerbert05}
has become an important tool for simulations of an increasingly wider range of problems in geology, 
material science, chemistry, and biology. 
{\em Ab initio} molecular dynamics, where the atomic positions move along
classical trajectories,
can be categorized in two major groups: Lagrangian Car-Parrinello molecular dynamics 
and Born-Oppenheimer molecular dynamics
\cite{CLeforestier,RCar85,RBarnett91,RWentzcovitch91,MCPayne92,DMarx00,HBSchlegel01,PPulay04,JMHerbert05}.
The tremendous success of Car-Parinello molecular dynamics, invented two decades ago
\cite{RCar85}, is based on its low computational cost combined with the
conserved Lagrangian properties of the dynamics. However, unless a Car-Parinello simulation
is performed carefully, it may yield results different from Born-Oppenheimer
molecular dynamics \cite{MCPayne92,DGibson95,DMarx00,PTangney02,JCGrossman04,ESchwegler04}. In Born-Oppenheimer molecular
dynamics the atomic positions are propagated by forces that are calculated
at the self-consistent electronic ground state for each instantaneous 
arrangement of the ions.  
Born-Oppenheimer molecular dynamics is computationally expensive
compared to Car-Parinello dynamics because of the requirement to reach 
a self-consistent field (SCF) solution in each time-step. 
However, the number of SCF cycles and thus the computational cost 
can be strongly reduced by using an initial guess
for the electronic degrees of freedom $\rho(t_{n+1})$ (here represented by the electron
density), which is given by an extrapolation from previous time steps
\cite{MCPayne92,TAArias92,JMMillan99,PPulay04,CRaynaud04,JMHerbert05}. 
The electronic extrapolation scheme, combined with the SCF procedure, 
can be seen as an adiabatic propagation of the electronic degrees of freedom, where
\begin{equation}\label{EXTRP}
\rho(t_{n+1}) = {\rm SCF}[\rho(t_n),\rho(t_{n-1}), \ldots]~.
\end{equation}
Unfortunately, this approach has a fundamental problem.
{\it Because of the non-linear and irreversible SCF procedure, which in practice 
never is complete, the time-reversal symmetry of the electronic propagation is broken.}
This problem does not occur in Lagrangian Car-Parinello molecular dynamics
\cite{PTangney06}, where both the nuclear and
the electronic degrees of freedom can be propagated with time-reversible Verlet integrators \cite{LVerlet67}.
The main purpose of this letter is to show how an effective time-reversible
propagation of the electronic degrees of freedom is possible in Born-Oppenheimer
molecular dynamics, despite an irreversible and approximate SCF procedure. 

Computational schemes using time-reversible integrators
give an energy meandering around a constant value that does not drift with time.  
This stability follows from the time-reversal symmetry, which excludes a steady 
increase or decrease of the energy (over the Poincare' time) for a periodic motion.
Because of the broken time-reversal symmetry in the adiabatic propagation of the 
electron density in Eq.\ (\ref{EXTRP}), a small but systematic energy drift 
occurs in the evolution of a microcanonical ensemble, with an accumulating phase 
space error. The error can be systematically reduced by
improving the SCF convergence or be removed completely 
by using an initial guess in the SCF procedure that is independent 
of previous time steps \cite{PPulay04,JMHerbert05}.  Both these remedies are 
computationally expensive and using an initial guess that is independent from
previous time steps often require a significantly increased number of SCF cycles.

\begin{figure}[t]
\resizebox*{3.0in}{!}{\includegraphics[angle=0]{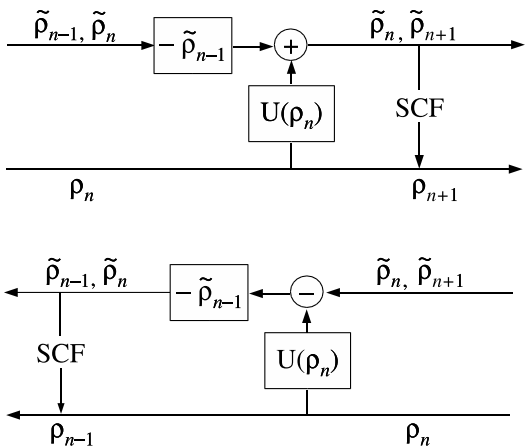}}
\caption{\label{Filter} 
The principle behind the time-reversible lossless filter process for propagation of the
electronic density.  In the forward filter process (upper part),
${\widetilde \rho}_{n+1} =  - {\widetilde \rho}_{n-1} + U(\rho_n)$ and
in the lossless backward reconstruction (lower part),
$-{\widetilde \rho}_{n-1} = {\widetilde \rho}_{n+1} - U(\rho_n)$. 
The dual propagation with the auxiliary density ${\widetilde \rho}_n$ 
allows a perfect reconstruction of the self-consistent Born-Oppenheimer 
density $\rho_n$, despite an irreversible, incomplete and approximate SCF procedure. }
\end{figure}

The basic principles for the time-reversible lossless integration of the electronic
degrees of freedom is described here in terms of the propagation of 
the electron density $\rho(t)$.  However, time-reversible 
Born-Oppenheimer trajectories can be constructed by replacing the density
by other parameters governing the electronic degrees of freedom, 
such as the effective single-particle potential, Hamiltonian, density matrix,
or wavefunctions. Our approach is therefore general and applicable to a large number of electronic 
structure schemes based on self-consistent Hartree-Fock or density functional theory. 

A reversible propagation of the electron density $\rho(t)$ 
in finite time steps of $\delta t$ ($t_n = t_0 + n\delta t$) can be constructed from a lossless filter
process analogous to, for example, the lossless wavelet transform used in data compression 
\cite{ARCalderbank98}.  The principle of the process, which is the key result in
this letter, is shown in the upper part of Fig.\ \ref{Filter}.  The scheme is divided into two
channels; the upper channel, with an approximate auxiliary density (denoted by a tilde), 
${\widetilde \rho}_{n+1} \equiv {\widetilde \rho}(t_{n+1})$,
is used as an initial guess for the Born-Oppenheimer density,
$\rho_{n+1}$, in the lower channel. The Born-Oppenheimer density is given through 
the non-linear, in practice incomplete, and numerically lossy SCF procedure,
\begin{equation}
\rho_{n+1} = {\rm SCF}[{\widetilde \rho}_{n+1}].
\end{equation}
The function $U(\rho_n)$, which is allowed to be numerically approximate and 
without a unique inverse, is an update filter for the propagation of the auxiliary density,
\begin{equation}\label{First}
{\widetilde \rho}_{n+1} = U(\rho_n) - {\widetilde \rho}_{n-1}.
\end{equation}
It is easy to see that the filter process is perfectly lossless and reversible by running
the process backwards in time with the $\oplus$ sign replaced by a $\ominus$
sign, as shown in the lower part of Fig. \ref{Filter}. The scheme is therefore a bijective
map which allows perfect reconstruction of $\rho_n$ backwards in time, despite the fact 
that the SCF procedure by itself is an irreversible lossy transform. 

The auxiliary density ${\widetilde \rho_{n+1}}$ in Eq.\ (\ref{First}) will be close 
to the self-consistent Born-Oppenheimer density $\rho_{n+1}$ if the lossless filter process in Fig.\ \ref{Filter} 
approximates the time-reversible adiabatic evolution of the density on the Born-Oppenheimer 
potential energy surface. This reduces the number of SCF cycles necessary
to reach the new self-consistent density.  One way to achieve this is to construct the update
filter $U(\rho_n)$ in Eq.\ (\ref{First}) from the time-reversible Verlet
integrator \cite{LVerlet67} such that
\begin{equation} 
{\widetilde \rho}_{n+1} = [2\rho_n + \delta t^2{\ddot \rho}_n] - {\widetilde \rho}_{n-1}.
\end{equation}
This integrator fulfills time-reversal symmetry since it remains the same
if we switch the sign of $\delta t$ and thus interchange ${\widetilde \rho}_{n-1}$
with ${\widetilde \rho}_{n+1}$. Note that perfect lossless reconstruction
is a necessary, but not a sufficient, condition for time-reversal 
symmetry.  The simplest time-reversible approximation of the second order time derivative, 
${\ddot \rho}_n = \partial^2 \rho_n / \partial t^2$,
is to set it equal to zero.  In this case the lossless propagation of the auxiliary density is 
\begin{equation}\label{Order1}
{\widetilde \rho}_{n+1} = 2\rho_n - {\widetilde \rho}_{n-1}.
\end{equation}
This surprisingly simple difference approximation fulfills time-reversal
symmetry, allows a perfect reconstruction backwards in time, and approximates
the propagation of the density with a modified linear interpolation from two
previous time steps.  It also avoids an unstable exponential error growth since the 
characteristic equation (with $\rho_n$ replaced by ${\widetilde \rho}_n$)
has no roots outside the unit circle. If the auxiliary density
${\widetilde \rho}_{n-1}$ in Eq.\ (\ref{Order1}) is replaced by 
the self-consistent Born-Oppenheimer density $\rho_{n-1}$ the propagation scheme is
identical to a linear interpolation.  Below we demonstrate how such a modification 
affects phase-space conservation and the energy drift.

A quite general scheme for constructing efficient time-reversible integrators for
the approximate adiabatic propagation of the electronic degrees of freedom is given
by a least square fit of the ansatz
\begin{equation}\label{TRAnsatz}
{\widetilde \rho}(t) = \sum_{m=0}^M a_m t^{2m} - {\widetilde \rho}(-t), 
\end{equation}
to Born-Oppenheimer densities $\rho(t_n)$ at $N$ successive time steps. A similar least square 
approximation, but without the constraint of time-reversibility, was recently
proposed by Pulay and Fogarasi for the extrapolation of the single-particle Hamiltonian in
a highly efficient Fock matrix dynamics (FMD) method \cite{PPulay04,JMHerbert05}.
Often only 2-3 SCF cycles are necessary in their scheme, but because of the broken time-reversal
symmetry a small but systematic energy drift occurs.

By choosing different numbers of fitted values $N$ we can calculate the expansion coefficients
$a_m$ in Eq.\ (\ref{TRAnsatz}) and express them in terms of previous Born-Oppenheimer densities. 
For example, a least square fit using the ansatz in Eq.\ (\ref{TRAnsatz}), with
$M=1$ for 6 previous densities, leads to the time-reversible approximation
\begin{equation}\label{6step}
{\widetilde \rho}_{n+1} =
\frac{1}{13}[30(\rho_n+\rho_{n-4}) - 3(\rho_{n-1}+\rho_{n-3}) - 28\rho_{n-2}] 
- {\widetilde \rho}_{n-5}.
\end{equation}
Note that the least square ansatz in Eq.\ (\ref{TRAnsatz}) also can be extended
to negative time exponents that are even.

\begin{figure}[t]
\resizebox*{3.0in}{!}{\includegraphics[angle=00]{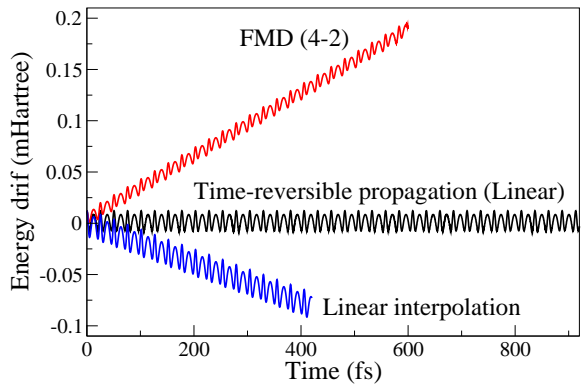}}
\caption{\label{Energy_Cons}
The fluctuations in the total energy as a function of time for a F$_2$ molecule 
using Hartree-Fock theory with a Gaussian basis set (RHF/3-21G).
The time-reversible propagation based on Eq.\ (\ref{Order1}), 
with the density replaced by the effective single-particle Hamiltonian,
is compared to the energy drift using the corresponding linear 
interpolation from previous time steps. The (4-2) Fock matrix dynamics (FDM)
scheme by Pulay and Fogarasi is shown as a comparison.
The time step was chosen to $\delta t = 0.25$ fs with 2 SCF/step.}
\end{figure}

\begin{figure}[t]
\resizebox*{3.0in}{!}{\includegraphics[angle=00]{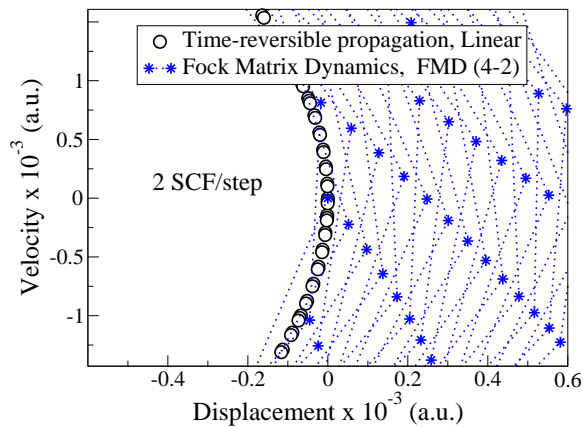}}
\caption{\label{Phase_Cons}
Detail of the phase space during 500 fs for one of the atoms in the $F_2$ molecule
(Hartree-Fock theory with a Gaussian basis set RHF/3-21G, 2 SCF/step).
The time-reversible propagation preserves the phase space whereas
the non-time-reversible dynamics has a small but noticeable drift.}
\end{figure}

To demonstrate the time-reversible lossless Born-Oppenheimer molecular dynamics we have implemented
the scheme in MondoSCF \cite{MondoSCF}, a suite of programs using Gaussian basis sets for 
electronic structure calculations based on self-consistent Hartree-Fock or
density functional theory.  The density in the time-reversible propagation 
in Eqs.\ (\ref{Order1}) and (\ref{TRAnsatz}) has been replaced by 
the effective single-particle Hamiltonian. The number of SCF cycles has been 
measured in the number of constructions of Hamiltonians, which is the 
most time consuming step.  

Figure \ref{Energy_Cons} shows the total energy as a function of time
for an $F_2$ molecule. The modification of the time-reversible linear propagation 
in Eq.\ (\ref{Order1}) to a linear interpolation leads to a systematic drift in 
the energy.  Using more SCF cycles per time step reduces
the energy drift, but it never really disappears. 
As a comparison we also show the (4-2) Fock matrix dynamics (FMD) scheme 
by Pulay and Fogarasi \cite{PPulay04}, based on a second
order polynomial least square fit using four previous data points.
This scheme gives, in principle, a more accurate extrapolation, which
is noticed in a smaller amplitude of the energy oscillations. However,
because of the broken time-reversal symmetry there is a systematic drift
in the energy.  For the time-reversible linear integrator in Eq.\
(\ref{Order1}), using only 2 SCF iterations per time step, any energy drift 
was less than $10^{-8}$ Hartree/ps.  The phase space is also conserved with 
the time-reversible integration, as illustrated in Fig.\ (\ref{Phase_Cons}).
Thus, even with an incomplete SCF convergence time-reversibility is correctly preserved.

\begin{figure}[t]
\resizebox*{3.0in}{!}{\includegraphics[angle=00]{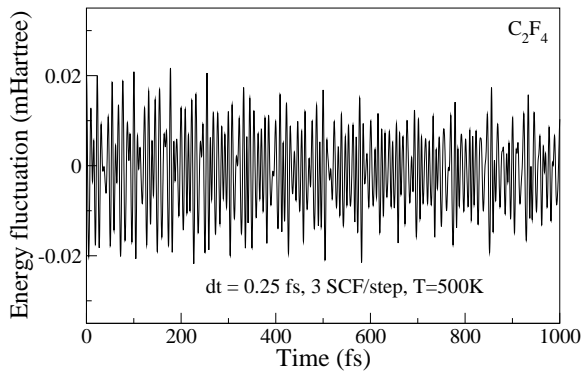}}
\caption{\label{EFluct} 
The energy fluctuations around the average energy as a function of time
for a C$_2$F$_4$ molecule at a temperature $T \approx 500$K
(Hartree-Fock theory with a Gaussian basis set RHF/3-21G). 
Three SCF cycles per time step of length $\delta t = 0.25$ fs were used.}
\end{figure}

Though unproven, all time-reversible propagators we have found so far have 
characteristic equations with all their roots on the unit circle. 
Because of the perfect lossless reconstruction, any error 
that occurs in the calculations will propagate throughout the simulation. 
This leads to a random noise that increases with time. Because of this noise
the auxiliary density ${\widetilde \rho_n}$ slowly moves away from the self-consistent
solution. This means that it is not possible to simply take
long time steps using only 1 SCF cycle per time step.
An increased number of SCF cycles (or shorter time steps) is in general 
necessary to reach a sufficiently accurate Born-Oppenheimer density for
longer simulation times.  Figure \ref{EFluct} shows the fluctuations in the total energy
for a C$_2$F$_4$ molecule during 1 ps of simulation time at a temperature $T \approx 500$K. 
The extrapolation scheme in Eq.\ (\ref{6step}) was used and 3 SCF cycles per time step was 
applied using Pulay's direct inversion in the iterative subspace (DIIS) algorithm to
accelerate the convergence \cite{PPulay80,PPulay82}. 

The ability of a perfect reconstruction of the dynamical data backwards in time 
is kept also using approximate arithmetics, thanks to the lossless bijective filter
process illustrated in Fig.\ {\ref{Filter}. The lossless property is therefore kept 
also when small elements below some drop tolerance are set to zero
in the SCF optimization of $\rho_n$ and in the update function $U(\rho_n)$. 
This is potentially important in the study of very large systems using, for example, 
linear scaling electronic structure methods \cite{SGoedecker99} 
that in general require a lower numerical accuracy compared to conventional schemes.


The main result in this letter is that we have shown that an effective
time-reversible propagation of the electronic degrees of freedom is possible
in self-consistent Born-Oppenheimer molecular dynamics, despite the irreversible, 
non-linear, and in practice always approximate SCF procedure. This ability
may also open the way for the development of higher order symplecitc 
integrators \cite{RRuth83,Eforest90,BJLeimkuhler94,DIOkunbor95,GJMartyna95} in
{\em ab initio} molecular dynamics,

In summary, we have presented a scheme for time-reversible Born-Oppenheimer molecular
dynamics that combines a low computational cost with a physically
correct time-reversible representation of the dynamics. 
The principle is based on a lossless filter
integration of the electronic degrees of freedom, where a dual propagation with an auxiliary
density allows a perfect reconstruction backwards in time of the self-consistent 
Born-Oppenheimer density. 
Time-reversal symmetry removes a systematic long term energy drift for a
microcanonical ensemble. This energy stabilization,
combined with the adiabatic propagation of an approximate Born-Oppenheimer 
electron density, strongly reduces the number of necessary self-consistency cycles,
while maintaining an accurate physical description. 

This work was performed under the auspices of the US Department of Energy,
Office of Science and LANL LDRD/ER program. A.N. acknowledge support form
the Swedish Research Council (VR).


\begin{thebibliography}{31}
\expandafter\ifx\csname natexlab\endcsname\relax\def\natexlab#1{#1}\fi
\expandafter\ifx\csname bibnamefont\endcsname\relax
  \def\bibnamefont#1{#1}\fi
\expandafter\ifx\csname bibfnamefont\endcsname\relax
  \def\bibfnamefont#1{#1}\fi
\expandafter\ifx\csname citenamefont\endcsname\relax
  \def\citenamefont#1{#1}\fi
\expandafter\ifx\csname url\endcsname\relax
  \def\url#1{\texttt{#1}}\fi
\expandafter\ifx\csname urlprefix\endcsname\relax\def\urlprefix{URL }\fi
\providecommand{\bibinfo}[2]{#2}
\providecommand{\eprint}[2][]{\url{#2}}
                                                                                                                                            
\bibitem[{\citenamefont{Roothaan}(1951)}]{Roothaan}
\bibinfo{author}{\bibfnamefont{C.~C.~J.} \bibnamefont{Roothaan}},
  \bibinfo{journal}{Rev. Mod. Phys.} \textbf{\bibinfo{volume}{23}},
  \bibinfo{pages}{69} (\bibinfo{year}{1951}).
                                                                                                                                            
\bibitem[{\citenamefont{Hohenberg and Kohn}(1964)}]{hohen}
\bibinfo{author}{\bibfnamefont{P.}~\bibnamefont{Hohenberg}} \bibnamefont{and}
  \bibinfo{author}{\bibfnamefont{W.}~\bibnamefont{Kohn}},
  \bibinfo{journal}{Phys. Rev.} \textbf{\bibinfo{volume}{136}},
  \bibinfo{pages}{B:864} (\bibinfo{year}{1964}).
                                                                                                                                            
\bibitem[{\citenamefont{Kohn and Sham}(1965)}]{KohnSham65}
\bibinfo{author}{\bibfnamefont{W.}~\bibnamefont{Kohn}} \bibnamefont{and}
  \bibinfo{author}{\bibfnamefont{L.~J.} \bibnamefont{Sham}},
  \bibinfo{journal}{Phys. Rev.} \textbf{\bibinfo{volume}{140}},
  \bibinfo{pages}{1133} (\bibinfo{year}{1965}).
                                                                                                                                            
\bibitem[{\citenamefont{Car and Parrinello}(1985)}]{RCar85}
\bibinfo{author}{\bibfnamefont{R.}~\bibnamefont{Car}} \bibnamefont{and}
  \bibinfo{author}{\bibfnamefont{M.}~\bibnamefont{Parrinello}},
  \bibinfo{journal}{Phys.\ Rev.\ Lett.} \textbf{\bibinfo{volume}{55}},
  \bibinfo{pages}{2471} (\bibinfo{year}{1985}).
                                                                                                                                            
\bibitem[{\citenamefont{Payne et~al.}(1992)\citenamefont{Payne, Teter, Allan,
  Arias, and Joannopoulos}}]{MCPayne92}
\bibinfo{author}{\bibfnamefont{M.~C.} \bibnamefont{Payne}},
  \bibinfo{author}{\bibfnamefont{M.~P.} \bibnamefont{Teter}},
  \bibinfo{author}{\bibfnamefont{D.~C.} \bibnamefont{Allan}},
  \bibinfo{author}{\bibfnamefont{T.~A.} \bibnamefont{Arias}}, \bibnamefont{and}
  \bibinfo{author}{\bibfnamefont{J.~D.} \bibnamefont{Joannopoulos}},
  \bibinfo{journal}{Rev. Mod. Phys.} \textbf{\bibinfo{volume}{64}},
  \bibinfo{pages}{1045} (\bibinfo{year}{1992}).
                                                                                                                                            
\bibitem[{\citenamefont{Marx and Hutter}(2000)}]{DMarx00}
\bibinfo{author}{\bibfnamefont{D.}~\bibnamefont{Marx}} \bibnamefont{and}
  \bibinfo{author}{\bibfnamefont{J.}~\bibnamefont{Hutter}},
  \emph{\bibinfo{title}{Modern Methods and Algorithms of Quantum Chemistry}}
  (\bibinfo{publisher}{ed. J. Grotendorst}, \bibinfo{address}{John von Neumann
  Institute for Computing, J\"ulich, Germany}, \bibinfo{year}{2000}),
  \bibinfo{edition}{2nd} ed.
                                                                                                                                            
\bibitem[{\citenamefont{Schlegel et~al.}(2001)\citenamefont{Schlegel, Millam,
  Iyengar, Voth, Daniels, Scusseria, and Frisch}}]{HBSchlegel01}
\bibinfo{author}{\bibfnamefont{H.~B.} \bibnamefont{Schlegel}},
  \bibinfo{author}{\bibfnamefont{J.~M.} \bibnamefont{Millam}},
  \bibinfo{author}{\bibfnamefont{S.~S.} \bibnamefont{Iyengar}},
  \bibinfo{author}{\bibfnamefont{G.~A.} \bibnamefont{Voth}},
  \bibinfo{author}{\bibfnamefont{A.~D.} \bibnamefont{Daniels}},
  \bibinfo{author}{\bibfnamefont{G.}~\bibnamefont{Scusseria}},
  \bibnamefont{and} \bibinfo{author}{\bibfnamefont{M.~J.}
  \bibnamefont{Frisch}}, \bibinfo{journal}{J. Chem. Phys.}
  \textbf{\bibinfo{volume}{114}}, \bibinfo{pages}{9758} (\bibinfo{year}{2001}).
                                                                                                                                            
\bibitem[{\citenamefont{Pulay and Fogarasi}(2004)}]{PPulay04}
\bibinfo{author}{\bibfnamefont{P.}~\bibnamefont{Pulay}} \bibnamefont{and}
  \bibinfo{author}{\bibfnamefont{G.}~\bibnamefont{Fogarasi}},
  \bibinfo{journal}{Chem. Phys. Lett.} \textbf{\bibinfo{volume}{386}},
  \bibinfo{pages}{272} (\bibinfo{year}{2004}).
                                                                                                                                            
\bibitem[{\citenamefont{Herbert and Head-Gordon}(2005)}]{JMHerbert05}
\bibinfo{author}{\bibfnamefont{J.}~\bibnamefont{Herbert}} \bibnamefont{and}
  \bibinfo{author}{\bibfnamefont{M.}~\bibnamefont{Head-Gordon}},
  \bibinfo{journal}{Phys. Chem. Chem. Phys.} \textbf{\bibinfo{volume}{7}},
  \bibinfo{pages}{3269} (\bibinfo{year}{2005}).
                                                                                                                                            
\bibitem[{\citenamefont{Leforestier}(1978)}]{CLeforestier}
\bibinfo{author}{\bibfnamefont{C.}~\bibnamefont{Leforestier}},
  \bibinfo{journal}{J. Chem. Phys.} \textbf{\bibinfo{volume}{68}},
  \bibinfo{pages}{4406} (\bibinfo{year}{1978}).
                                                                                                                                            
\bibitem[{\citenamefont{Barnett et~al.}(1991)\citenamefont{Barnett, Landman,
  Nitzan, and Rajagopal}}]{RBarnett91}
\bibinfo{author}{\bibfnamefont{R.~N.} \bibnamefont{Barnett}},
  \bibinfo{author}{\bibfnamefont{U.}~\bibnamefont{Landman}},
  \bibinfo{author}{\bibfnamefont{A.}~\bibnamefont{Nitzan}}, \bibnamefont{and}
  \bibinfo{author}{\bibfnamefont{G.}~\bibnamefont{Rajagopal}},
  \bibinfo{journal}{J. Chem. Phys.} \textbf{\bibinfo{volume}{94}},
  \bibinfo{pages}{608} (\bibinfo{year}{1991}).
                                                                                                                                            
\bibitem[{\citenamefont{Wentzcovitch and Martins}(1991)}]{RWentzcovitch91}
\bibinfo{author}{\bibfnamefont{R.~M.} \bibnamefont{Wentzcovitch}}
  \bibnamefont{and} \bibinfo{author}{\bibfnamefont{J.~L.}
  \bibnamefont{Martins}}, \bibinfo{journal}{Solid. State. Commun.}
  \textbf{\bibinfo{volume}{78}}, \bibinfo{pages}{831} (\bibinfo{year}{1991}).
                                                                                                                                            
\bibitem[{\citenamefont{Tangney and Scandolo}(2002)}]{PTangney02}
\bibinfo{author}{\bibfnamefont{P.}~\bibnamefont{Tangney}} \bibnamefont{and}
  \bibinfo{author}{\bibfnamefont{S.}~\bibnamefont{Scandolo}},
  \bibinfo{journal}{J. Chem. Phys.} \textbf{\bibinfo{volume}{116}},
  \bibinfo{pages}{14} (\bibinfo{year}{2002}).
                                                                                                                                            
\bibitem[{\citenamefont{Grossman et~al.}(2004)\citenamefont{Grossman,
  Schwegler, Draeger, Gygi, and Galli}}]{JCGrossman04}
\bibinfo{author}{\bibfnamefont{J.~C.} \bibnamefont{Grossman}},
  \bibinfo{author}{\bibfnamefont{E.}~\bibnamefont{Schwegler}},
  \bibinfo{author}{\bibfnamefont{E.~W.} \bibnamefont{Draeger}},
  \bibinfo{author}{\bibfnamefont{F.}~\bibnamefont{Gygi}}, \bibnamefont{and}
  \bibinfo{author}{\bibfnamefont{G.}~\bibnamefont{Galli}}, \bibinfo{journal}{J.
  Chem. Phys.} \textbf{\bibinfo{volume}{120}}, \bibinfo{pages}{300}
  (\bibinfo{year}{2004}).
                                                                                                                                            
\bibitem[{\citenamefont{Schwegler et~al.}(2004)\citenamefont{Schwegler,
  Grossman, Gygi, and Galli}}]{ESchwegler04}
\bibinfo{author}{\bibfnamefont{E.}~\bibnamefont{Schwegler}},
  \bibinfo{author}{\bibfnamefont{J.~C.} \bibnamefont{Grossman}},
  \bibinfo{author}{\bibfnamefont{F.}~\bibnamefont{Gygi}}, \bibnamefont{and}
  \bibinfo{author}{\bibfnamefont{G.}~\bibnamefont{Galli}}, \bibinfo{journal}{J.
  Chem. Phys.} \textbf{\bibinfo{volume}{121}}, \bibinfo{pages}{5400}
  (\bibinfo{year}{2004}).
                                                                                                                                            
\bibitem[{\citenamefont{Gibson et~al.}(1995)\citenamefont{Gibson, Ionova, and
  Carter}}]{DGibson95}
\bibinfo{author}{\bibfnamefont{D.~A.} \bibnamefont{Gibson}},
  \bibinfo{author}{\bibfnamefont{I.~V.} \bibnamefont{Ionova}},
  \bibnamefont{and} \bibinfo{author}{\bibfnamefont{E.}~\bibnamefont{Carter}},
  \bibinfo{journal}{Chem. Phys. Lett.} \textbf{\bibinfo{volume}{240}},
  \bibinfo{pages}{261} (\bibinfo{year}{1995}).
                                                                                                                                            
\bibitem[{\citenamefont{Arias et~al.}(1992)\citenamefont{Arias, Payne, and
  Joannopoulos}}]{TAArias92}
\bibinfo{author}{\bibfnamefont{T.}~\bibnamefont{Arias}},
  \bibinfo{author}{\bibfnamefont{M.}~\bibnamefont{Payne}}, \bibnamefont{and}
  \bibinfo{author}{\bibfnamefont{J.}~\bibnamefont{Joannopoulos}},
  \bibinfo{journal}{Phys. Rev. Lett.} \textbf{\bibinfo{volume}{69}},
  \bibinfo{pages}{1077} (\bibinfo{year}{1992}).
                                                                                                                                            
\bibitem[{\citenamefont{Millan et~al.}(1999)\citenamefont{Millan, Bakken, Chen,
  Hase, and Schlegel}}]{JMMillan99}
\bibinfo{author}{\bibfnamefont{J.}~\bibnamefont{Millan}},
  \bibinfo{author}{\bibfnamefont{V.}~\bibnamefont{Bakken}},
  \bibinfo{author}{\bibfnamefont{W.}~\bibnamefont{Chen}},
  \bibinfo{author}{\bibfnamefont{L.}~\bibnamefont{Hase}}, \bibnamefont{and}
  \bibinfo{author}{\bibfnamefont{H.~B.} \bibnamefont{Schlegel}},
  \bibinfo{journal}{J. Chem. Phys.} \textbf{\bibinfo{volume}{111}},
  \bibinfo{pages}{3800} (\bibinfo{year}{1999}).
                                                                                                                                            
\bibitem[{\citenamefont{Raynaud et~al.}(2004)\citenamefont{Raynaud, Maron,
  Daudey, and Jolibois}}]{CRaynaud04}
\bibinfo{author}{\bibfnamefont{C.}~\bibnamefont{Raynaud}},
  \bibinfo{author}{\bibfnamefont{L.}~\bibnamefont{Maron}},
  \bibinfo{author}{\bibfnamefont{J.-P.} \bibnamefont{Daudey}},
  \bibnamefont{and} \bibinfo{author}{\bibfnamefont{F.}~\bibnamefont{Jolibois}},
  \bibinfo{journal}{Phys. Chem. Phys.} \textbf{\bibinfo{volume}{6}},
  \bibinfo{pages}{4226} (\bibinfo{year}{2004}).
                                                                                                                                            
\bibitem[{\citenamefont{Tangney}(2006)}]{PTangney06}
\bibinfo{author}{\bibfnamefont{P.}~\bibnamefont{Tangney}}, \bibinfo{journal}{J.
  Chem. Phys.} \textbf{\bibinfo{volume}{124}}, \bibinfo{pages}{44111}
  (\bibinfo{year}{2006}).
                                                                                                                                            
\bibitem[{\citenamefont{Verlet}(1967)}]{LVerlet67}
\bibinfo{author}{\bibfnamefont{L.}~\bibnamefont{Verlet}},
  \bibinfo{journal}{Phys. Rev.} \textbf{\bibinfo{volume}{159}},
  \bibinfo{pages}{98} (\bibinfo{year}{1967}).
                                                                                                                                            
\bibitem[{\citenamefont{Calderbank et~al.}(1998)\citenamefont{Calderbank,
  Daubechies, Sweldens, and Yeo}}]{ARCalderbank98}
\bibinfo{author}{\bibfnamefont{A.~R.} \bibnamefont{Calderbank}},
  \bibinfo{author}{\bibfnamefont{I.}~\bibnamefont{Daubechies}},
  \bibinfo{author}{\bibfnamefont{W.}~\bibnamefont{Sweldens}}, \bibnamefont{and}
  \bibinfo{author}{\bibfnamefont{B.~L.} \bibnamefont{Yeo}},
  \bibinfo{journal}{Appl. Comput. Harmon. Anal.} \textbf{\bibinfo{volume}{5}},
  \bibinfo{pages}{332} (\bibinfo{year}{1998}).

\bibitem[{\citenamefont{Challacombe et~al.}(2001)\citenamefont{Challacombe,
  Schwegler, Tymczak, Gan, Nemeth, Weber, Niklasson, and Henkelman}}]{MondoSCF}
\bibinfo{author}{\bibfnamefont{M.}~\bibnamefont{Challacombe}},
  \bibinfo{author}{\bibfnamefont{E.}~\bibnamefont{Schwegler}},
  \bibinfo{author}{\bibfnamefont{C.~J.} \bibnamefont{Tymczak}},
  \bibinfo{author}{\bibfnamefont{C.~K.} \bibnamefont{Gan}},
  \bibinfo{author}{\bibfnamefont{K.}~\bibnamefont{Nemeth}},
  \bibinfo{author}{\bibfnamefont{V.}~\bibnamefont{Weber}},
  \bibinfo{author}{\bibfnamefont{A.~M.~N.} \bibnamefont{Niklasson}},
  \bibnamefont{and}
  \bibinfo{author}{\bibfnamefont{G.}~\bibnamefont{Henkelman}},
  \emph{\bibinfo{title}{{\sc MondoSCF} v1.0$\alpha$9}} (\bibinfo{year}{2001}),
  \bibinfo{note}{\mbox{L}os Alamos National Laboratory (LA-CC 01-2), Copyright
  University of California.},
  \urlprefix\url{http://www.t12.lanl.gov/home/mchalla/}.
                                                                                                                                            
\bibitem[{\citenamefont{Pulay}(1980)}]{PPulay80}
\bibinfo{author}{\bibfnamefont{P.}~\bibnamefont{Pulay}},
  \bibinfo{journal}{Chem. Phys. Let.} \textbf{\bibinfo{volume}{73}},
  \bibinfo{pages}{393} (\bibinfo{year}{1980}).
                                                                                                                                            
\bibitem[{\citenamefont{Pulay}(1982)}]{PPulay82}
\bibinfo{author}{\bibfnamefont{P.}~\bibnamefont{Pulay}}, \bibinfo{journal}{J.
  Comput. Chem.} \textbf{\bibinfo{volume}{3}}, \bibinfo{pages}{556}
  (\bibinfo{year}{1982}).
                                                                                                                                            
\bibitem[{\citenamefont{Goedecker}(1999)}]{SGoedecker99}
\bibinfo{author}{\bibfnamefont{S.}~\bibnamefont{Goedecker}},
  \bibinfo{journal}{Rev. Mod. Phys.} \textbf{\bibinfo{volume}{71}},
  \bibinfo{pages}{1085} (\bibinfo{year}{1999}).
                                                                                                                                            
\bibitem[{\citenamefont{Ruth}(1983)}]{RRuth83}
\bibinfo{author}{\bibfnamefont{R.}~\bibnamefont{Ruth}}, \bibinfo{journal}{IEEE
  Trans. Nucl. Sci.} \textbf{\bibinfo{volume}{30}}, \bibinfo{pages}{2669}
  (\bibinfo{year}{1983}).
                                                                                                                                            
\bibitem[{\citenamefont{Forest and Ruth}(1990)}]{Eforest90}
\bibinfo{author}{\bibfnamefont{E.}~\bibnamefont{Forest}} \bibnamefont{and}
  \bibinfo{author}{\bibfnamefont{R.}~\bibnamefont{Ruth}},
  \bibinfo{journal}{Physica D} \textbf{\bibinfo{volume}{43}},
  \bibinfo{pages}{105} (\bibinfo{year}{1990}).
                                                                                                                                            
\bibitem[{\citenamefont{Leimkuhler and Skeel}(1994)}]{BJLeimkuhler94}
\bibinfo{author}{\bibfnamefont{B.~J.} \bibnamefont{Leimkuhler}}
  \bibnamefont{and} \bibinfo{author}{\bibfnamefont{R.~D.} \bibnamefont{Skeel}},
  \bibinfo{journal}{J. Comput. Phys.} \textbf{\bibinfo{volume}{112}},
  \bibinfo{pages}{117} (\bibinfo{year}{1994}).
                                                                                                                                            
\bibitem[{\citenamefont{Okunbor}(1995)}]{DIOkunbor95}
\bibinfo{author}{\bibfnamefont{D.~I.} \bibnamefont{Okunbor}},
  \bibinfo{journal}{J. Comput. Phys.} \textbf{\bibinfo{volume}{120}},
  \bibinfo{pages}{375} (\bibinfo{year}{1995}).
                                                                                                                                            
\bibitem[{\citenamefont{G.J.Martyna and Tuckerman}(1995)}]{GJMartyna95}
\bibinfo{author}{\bibnamefont{G.J.Martyna}} \bibnamefont{and}
  \bibinfo{author}{\bibfnamefont{M.}~\bibnamefont{Tuckerman}},
  \bibinfo{journal}{J. Chem. Phys.} \textbf{\bibinfo{volume}{102}},
  \bibinfo{pages}{8071} (\bibinfo{year}{1995}).
                                                                                                                                            
\end{thebibliography}

\end{document}